\documentstyle[twoside,epsf]{article}

\catcode`\@=11
\long\def\@makefntext#1{
\protect\noindent \hbox to 3.2pt {\hskip-.9pt  

$^{{\eightrm\@thefnmark}}$\hfil}#1\hfill}		

\def\thefootnote{\fnsymbol{footnote}}
\def\@makefnmark{\hbox to 0pt{$^{\@thefnmark}$\hss}}	

\def\ps@myheadings{\let\@mkboth\@gobbletwo
\def\@oddhead{\hbox{}
\rightmark\hfil\eightrm\thepage}   

\def\@oddfoot{}\def\@evenhead{\eightrm\thepage\hfil
\leftmark\hbox{}}\def\@evenfoot{}
\def\sectionmark##1{}\def\subsectionmark##1{}}

\oddsidemargin=\evensidemargin
\renewcommand{\thefootnote}{\fnsymbol{footnote}}

\newcounter{sectionc}\newcounter{subsectionc}\newcounter{subsubsectionc}
\renewcommand{\section}[1] {\vspace{12pt}\addtocounter{sectionc}{1} 

\setcounter{subsectionc}{0}\setcounter{subsubsectionc}{0}\noindent 

	{\tenbf\thesectionc. #1}\par\vspace{5pt}}
\renewcommand{\subsection}[1] {\vspace{12pt}\addtocounter{subsectionc}{1} 

	\setcounter{subsubsectionc}{0}\noindent 

	{\bf\thesectionc.\thesubsectionc. {\kern1pt \bfit #1}}\par\vspace{5pt}}
\renewcommand{\subsubsection}[1] {\vspace{12pt}\addtocounter{subsubsectionc}{1}
	\noindent{\tenrm\thesectionc.\thesubsectionc.\thesubsubsectionc.
	{\kern1pt \tenit #1}}\par\vspace{5pt}}
\newcommand{\nonumsection}[1] {\vspace{12pt}\noindent{\tenbf #1}
	\par\vspace{5pt}}

\newcounter{appendixc}
\newcounter{subappendixc}[appendixc]
\newcounter{subsubappendixc}[subappendixc]
\renewcommand{\thesubappendixc}{\Alph{appendixc}.\arabic{subappendixc}}
\renewcommand{\thesubsubappendixc}
	{\Alph{appendixc}.\arabic{subappendixc}.\arabic{subsubappendixc}}

\renewcommand{\appendix}[1] {\vspace{12pt}
        \refstepcounter{appendixc}
        \setcounter{figure}{0}
        \setcounter{table}{0}
        \setcounter{lemma}{0}
        \setcounter{theorem}{0}
        \setcounter{corollary}{0}
        \setcounter{definition}{0}
        \setcounter{equation}{0}
        \renewcommand{\thefigure}{\Alph{appendixc}.\arabic{figure}}
        \renewcommand{\thetable}{\Alph{appendixc}.\arabic{table}}
        \renewcommand{\theappendixc}{\Alph{appendixc}}
        \renewcommand{\thelemma}{\Alph{appendixc}.\arabic{lemma}}
        \renewcommand{\thetheorem}{\Alph{appendixc}.\arabic{theorem}}
        \renewcommand{\thedefinition}{\Alph{appendixc}.\arabic{definition}}
        \renewcommand{\thecorollary}{\Alph{appendixc}.\arabic{corollary}}
        \renewcommand{\theequation}{\Alph{appendixc}.\arabic{equation}}
        \noindent{\tenbf Appendix \theappendixc #1}\par\vspace{5pt}}
\newcommand{\subappendix}[1] {\vspace{12pt}
        \refstepcounter{subappendixc}
        \noindent{\bf Appendix \thesubappendixc. {\kern1pt \bfit #1}}
	\par\vspace{5pt}}
\newcommand{\subsubappendix}[1] {\vspace{12pt}
        \refstepcounter{subsubappendixc}
        \noindent{\rm Appendix \thesubsubappendixc. {\kern1pt \tenit #1}}
	\par\vspace{5pt}}

\newcommand{\textlineskip}{\baselineskip=13pt}
\newcommand{\smalllineskip}{\baselineskip=10pt}

\def\eightcirc{
\begin{picture}(0,0)
\put(4.4,1.8){\circle{6.5}}
\end{picture}}
\def\eightcopyright{\eightcirc\kern2.7pt\hbox{\eightrm c}}

\newcommand{\copyrightheading}[1]
	{\vspace*{-2.5cm}\smalllineskip{\flushleft
	{\footnotesize PAR--LPTHE 96--31}\\
	{\footnotesize hep-th/9609049}\\
	 }}

\def\abstracts#1#2#3{{
	\centering{\begin{minipage}{4.5in}\baselineskip=10pt\footnotesize
	\parindent=0pt #1\par 

	\parindent=15pt #2\par
	\parindent=15pt #3
	\end{minipage}}\par}}

\newcommand{\bibit}{\nineit}

\renewenvironment{thebibliography}[1]
	{\frenchspacing
	 \ninerm\baselineskip=11pt
	 \begin{list}{\arabic{enumi}.}
	{\usecounter{enumi}\setlength{\parsep}{0pt}
	 \setlength{\leftmargin 12.7pt}{\rightmargin 0pt} 
	 \setlength{\itemsep}{0pt} \settowidth
	{\labelwidth}{#1.}\sloppy}}{\end{list}}

\newcounter{itemlistc}
\newcounter{romanlistc}
\newcounter{alphlistc}
\newcounter{arabiclistc}

\newcommand{\fcaption}[1]{
        \refstepcounter{figure}
        \setbox\@tempboxa = \hbox{\footnotesize Fig.~\thefigure. #1}
        \ifdim \wd\@tempboxa > 5in
           {\begin{center}
        \parbox{5in}{\footnotesize\smalllineskip Fig.~\thefigure. #1}
            \end{center}}
        \else
             {\begin{center}
             {\footnotesize Fig.~\thefigure. #1}
              \end{center}}
        \fi}

\newcommand{\tcaption}[1]{
        \refstepcounter{table}
        \setbox\@tempboxa = \hbox{\footnotesize Table~\thetable. #1}
        \ifdim \wd\@tempboxa > 5in
           {\begin{center}
        \parbox{5in}{\footnotesize\smalllineskip Table~\thetable. #1}
            \end{center}}
        \else
             {\begin{center}
             {\footnotesize Table~\thetable. #1}
              \end{center}}
        \fi}

\def\@citex[#1]#2{\if@filesw\immediate\write\@auxout
	{\string\citation{#2}}\fi
\def\@citea{}\@cite{\@for\@citeb:=#2\do
	{\@citea\def\@citea{,}\@ifundefined
	{b@\@citeb}{{\bf ?}\@warning
	{Citation `\@citeb' on page \thepage \space undefined}}
	{\csname b@\@citeb\endcsname}}}{#1}}

\newif\if@cghi
\def\cite{\@cghitrue\@ifnextchar [{\@tempswatrue
	\@citex}{\@tempswafalse\@citex[]}}
\def\citelow{\@cghifalse\@ifnextchar [{\@tempswatrue
	\@citex}{\@tempswafalse\@citex[]}}
\def\@cite#1#2{{$\null^{#1}$\if@tempswa\typeout
	{IJCGA warning: optional citation argument 

	ignored: `#2'} \fi}}

\def\pmb#1{\setbox0=\hbox{#1}
	\kern-.025em\copy0\kern-\wd0
	\kern.05em\copy0\kern-\wd0
	\kern-.025em\raise.0433em\box0}

\def\fnt#1#2{\footnotetext{\kern-.3em
	{$^{\mbox{\scriptsize #1}}$}{#2}}}

\def\fpage#1{\begingroup
\voffset=.3in
\thispagestyle{empty}\begin{table}[b]\centerline{\footnotesize #1}
	\end{table}\endgroup}

\def\runninghead#1#2{\pagestyle{myheadings}
\markboth{{\protect\footnotesize\it{\quad #1}}\hfill}
{\hfill{\protect\footnotesize\it{#2\quad}}}}
\font\tenrm=cmr10
\font\tenit=cmti10 

\font\tenbf=cmbx10
\font\bfit=cmbxti10 at 10pt
\font\ninerm=cmr9
\font\nineit=cmti9

\font\eightrm=cmr8

\def\qed{\hbox{${\vcenter{\vbox{			
   \hrule height 0.4pt\hbox{\vrule width 0.4pt height 6pt
   \kern5pt\vrule width 0.4pt}\hrule height 0.4pt}}}$}}

\renewcommand{\thefootnote}{\fnsymbol{footnote}}	

\begin{document}

\runninghead{Mark de Wild Propitius}  
{Unraveling spontaneously broken abelian Chern-Simons theories}

\normalsize\textlineskip
\thispagestyle{empty}
\setcounter{page}{1}

\copyrightheading{}			

\vspace*{0.88truein}

\fpage{1}
\centerline{\bf UNRAVELING SPONTANEOUSLY BROKEN}
\vspace*{0.035truein}
\centerline{\bf ABELIAN CHERN-SIMONS THEORIES\footnote{Based on a 
talk given at `The workshop on low dimensional field theory', Telluride,
Colorado, USA, August  5 to 17, 1996.}}
\vspace*{0.37truein}
\centerline{\footnotesize Mark de Wild Propitius\footnote{e-mail:
mdwp@lpthe.jussieu.fr}}
\vspace*{0.015truein}
\centerline{\footnotesize\it 
Laboratoire de Physique Th\'eorique et Haute 
Energies\footnote{Laboratoire associ\'e No.\ 280 au CNRS}}
\baselineskip=10pt
\centerline{\footnotesize\it Universit\'e Pierre et Marie Curie - PARIS VI}
\baselineskip=10pt
\centerline{\footnotesize\it Universit\'e Denis Diderot - PARIS VII} 
\baselineskip=10pt
\centerline{\footnotesize\it 4 place Jussieu, Boite 126, 
Tour 16, 1$^{er}$ \'etage}
\baselineskip=10pt
\centerline{\footnotesize\it F-75252 Paris CEDEX 05, France}
\vspace*{10pt}
\centerline{\footnotesize August 1996}

\vspace*{0.21truein}
\abstracts{In this talk I describe recent work (hep-th/9606029) in which I 
classified all conceivable 2+1 dimensional Chern-Simons (CS) theories 
with continuous compact abelian gauge group or finite abelian 
gauge group. The CS theories with finite abelian gauge group
that can be obtained from the spontaneous breakdown of a CS theory
with gauge group the direct product of various compact $U(1)$ gauge groups
were also identified. Those that can not be reached in this way are 
actually the most interesting since they lead to nonabelian phenomena 
such as nonabelian braid statistics, Alice fluxes and Cheshire charges 
and quite generally lead to dualities with 2+1 dimensional theories with 
a nonabelian finite gauge group.}{}{}

\newcommand{\bea}{\begin{eqnarray}}
\newcommand{\eea}{\end{eqnarray}}
\newcommand{\DH}{D(H)}
\newcommand{\DW}{D^{\omega}(H)}
\newcommand{\im}{\imath}
\newcommand{\Z}{{\mbox{\bf Z}}}
\newcommand{\al}{\alpha}
\newcommand{\bt}{\beta}
\newcommand{\gm}{\gamma}
\newcommand{\dl}{\delta}
\newcommand{\zt}{\zeta}
\newcommand{\et}{\eta}
\newcommand{\th}{\theta}
\newcommand{\kp}{\kappa}
\newcommand{\lm}{\lambda}
\newcommand{\rh}{\rho}
\newcommand{\hl}{\hline}
\newcommand{\sg}{\sigma}
\newcommand{\ta}{\tau}
\newcommand{\ph}{\phi}
\newcommand{\phv}{\varphi}
\newcommand{\ch}{\chi}
\newcommand{\ps}{\Phi}
\newcommand{\ot}{\otimes}


\vspace*{1pt}\textlineskip	
\vspace*{-0.5pt}
\noindent

\textheight=7.8truein
\setcounter{footnote}{0}
\renewcommand{\thefootnote}{\alph{footnote}}

\section{Introduction}
\noindent
Ever since the pioneering work of Schonfeld and Deser 
et al.~\cite{schonfeld}, 
2+1 dimensional Chern-Simons (CS) theories have received
quite some attention.
The motivations to study these theories range from applications in 
knot theory to applications in condensed matter systems such as 
fractional quantum Hall liquids. 
In this talk, I will dicuss the implications of adding a CS term 
to 2+1 dimensional gauge theories spontaneously broken down to a 
finite residual gauge group by means of the Higgs mechanism. That is, 
the focus is on models governed by an action of the form
\bea                               \label{alg}
S &=& S_{\mbox{\scriptsize YMH} } + S_{\mbox{\scriptsize matter}} 
+ S_{\mbox{\scriptsize CS}} \, ,
\eea
where the Yang--Mills--Higgs action $S_{\rm{\mbox{\scriptsize YMH}}}$ 
describes the spontaneous breakdown of some continuous compact gauge 
group $G$ to a finite subgroup ${ H}$,  
$S_{\mbox{\scriptsize matter}}$ a conserved matter current   
coupled to the gauge fields and
$S_{\mbox{\scriptsize CS}}$ denotes the CS action~\cite{schonfeld}. 

The so--called discrete $H$ gauge theories describing the long distance 
physics of the models~(\ref{alg}) without CS term 
have been studied in 2+1 and 3+1 dimensional space time and are by 
now completely understood. 
For a recent review and detailed references, 
see Ref.~\cite{banff}. 
To sketch the main results, the spectrum features topological 
defects which in 2+1 dimensional space time appear as  
vortices carrying 
magnetic flux labeled by the elements of $H$. If 
$H$ is nonabelian, the vortices exhibit a nonabelian Aharonov-Bohm 
(AB) effect: upon braiding two vortices their fluxes affect each 
other through conjugation. The residual 
global gauge group $H$ also acts on the fluxes through
conjugation, so the different magnetic vortices
are labeled by the conjugacy classes of $H$. This is in a nutshell the physics
described by the Yang-Mills Higgs part $S_{\mbox{\scriptsize YMH} }$
of the action~(\ref{alg}). 
The matter fields coupled to the gauge fields in 
$S_{\mbox{\scriptsize matter}}$ form multiplets which 
transform irreducibly under ${ G}$. In the broken 
phase these branch to irreducible representations of the residual gauge 
group ${ H}$. So the matter fields introduce point charges in the broken  
phase labeled by the unitary irreducible representations (UIR's) $\Gamma$ of 
$H$. If such a charge encircles a magnetic flux $h \in H$, 
it also undergoes an AB effect: it  returns 
transformed by the matrix $\Gamma(h)$. Since all gauge fields 
are massive, the foregoing AB effects form the only long range interactions 
among the charges and vortices. 
The complete spectrum also features dyons obtained 
by composing the vortices and charges. These are labeled by the conjugacy 
classes of $H$ paired with a nontrivial centralizer 
representation~\cite{spm}. A breakthrough in the understanding 
of these models
was the observation~\cite{spm} that this spectrum of charges, 
vortices and dyons together with the spin, braiding and fusion properties 
of these particles 
is, in fact, fully described by the representation 
theory of the quasitriangular Hopf 
algebra $D(H)$ resulting~\cite{dpr} from Drinfeld's double 
construction applied to the algebra 
${\cal F}(H)$ of functions on ${ H}$.    

As has been argued 
in Ref.~\cite{spm1}, 
the presence of a CS term $S_{\mbox{\scriptsize CS}}$ for the broken gauge 
group $G$ in the action~(\ref{alg}) gives rise to additional 
AB interactions among the vortices which are completely 
encoded in a 3-cocycle $\omega \in H^3(H,U(1))$ 
for the residual finite gauge group $H$.
The related algebraic structure is the quasi-Hopf algebra $\DW$ 
being a deformation of $D(H)$ by this 3-cocycle $\omega$.  
In Ref.~\cite{spm1}, these general results were just explicitly illustrated  
by the the abelian CS Higgs model in which the (compact) gauge group
$G \simeq U(1)$ is broken down to a cyclic subgroup $H \simeq \Z_N$.
Here, I will summarize the results of my recent paper~\cite{spabcst}
in which this analysis was extended  to spontaneously 
broken abelian CS theories in full generality. 
I will be rather burlesque concerning references. 
For civilized referencing, the reader has to consult Ref.~\cite{spabcst}.
As for conventions, natural units 
in which $\hbar=c=1$ are employed throughout. I will 
exclusively work in 2+1 dimensional Minkowsky space with signature 
$(+,-,-)$. Spatial coordinates are denoted by $x^1$ and $x^2$ and 
the time coordinate by $x^0$. Greek indices run from 0 to 2, whereas spatial 
components are labeled by latin indices $\in 1,2$.   

\section{The models, their spectrum and the AB interactions \label{model}}

\noindent
Let us concentrate on the subset of models~(\ref{alg}) 
realizing  symmetry breaking schemes 
$
{ G} \simeq U(1)^k   \rightarrow   H 
$ 
with  $U(1)^k$ the direct product of $k$ compact $U(1)$ gauge 
groups and the finite subgroup $H \simeq 
\Z_{N^{(1)}} \times \cdots \times
\Z_{N^{(k)}} $ a direct  product of $k$ cyclic groups  
$\Z_{N^{(i)}}$ of order $N^{(i)}$. So, the Yang--Mills--Higgs 
part of the action~(\ref{alg}) contains $k$  
complex scalar Higgs fields $\Phi^{(i)}$ (with $i \in 1,2,\ldots,k$) 
carrying charge $N^{(i)}e^{(i)}$ with $e^{(i)}$ 
the coupling constant for the $i^{th}$ compact $U(1)$ gauge field 
$A_{\kappa}^{(i)}$, i.e.
\bea   \label{ymh}
{S}_{\mbox{\scriptsize YMH}} &=& \int d\,^3x \; \left( 
\sum_{i=1}^k\{-\frac{1}{4}F^{(i)\kappa\nu} F^{(i)}_{\kappa\nu} + 
({\cal D}^\kappa \Phi^{(i)})^*{\cal D}_\kappa \Phi^{(i)} - 
V(|\Phi^{(i)}|)\} \right) ,
\eea
with  ${\cal D}_\kappa \Phi^{(i)}= 
(\partial_{\kappa}+\im N^{(i)}e^{(i)} A_{\kappa}^{(i)})\Phi^{(i)}$  and 
$ F^{(i)}_{\kappa\nu} =  \partial_{\kappa} A_{\nu}^{(i)}
-\partial_{\nu}A_\kappa^{(i)}$. All $U(1)$  gauge groups are 
assumed to be broken down at the same energy scale $M_H = v \sqrt{2\lambda}$.
Hence,
$
V(|\Phi^{(i)}|) = \frac{\lambda}{4}(|\Phi^{(i)}|^2-v^2)^2$ with 
$\lambda, v > 0$. In the matter part of~(\ref{alg}), 
we then have $k$ conserved matter currents $j^{(i)}$
coupled to the gauge fields
\bea \label{ma}
{S}_{\mbox{\scriptsize matter}} &=& \int d\,^3x \; \left( 
-\sum_{i=1}^k j^{(i)\kappa}A^{(i)}_{\kappa} \right)  .
\label{j12mat}
\eea 
The matter charges $q^{(i)}$ introduced by the current $j^{(i)}$
are supposed to be multiples of $e^{(i)}$.  
Finally, the most general CS action for this theory is of the form
\bea    \label{csact}
S_{\mbox{\scriptsize CS}} & = & 
\int d\,^3x \; \left( \sum_{1 \leq i<j \leq k} 
\; \frac{\mu^{(i)}}{2} \epsilon^{\kappa\sigma\rho}
                              A^{(i)}_{\kappa} \partial_{\sigma} 
                              A^{(i)}_{\rho} \label{CSt1} +
  \frac{\mu^{(ij)}}{2} \epsilon^{\kappa\sigma\rho}
                              A^{(i)}_{\kappa} \partial_{\sigma} 
                              A^{(j)}_{\rho} \right) ,     
\eea
with $\mu^{(i)}$ and 
$\mu^{(ij)}$ the topological masses and $\epsilon^{\kappa\sigma\rho}$ 
the three dimensional anti-symmetric Levi-Civita tensor normalized
such that $\epsilon^{012}=1$. Hence, there are  $k$ distinct CS 
terms $(i)$ describing self couplings of the  $U(1)$ gauge fields.
In addition, there are  $\frac{1}{2}k(k-1)$ 
distinct CS terms $(ij)$ 
establishing  pairwise couplings between different $U(1)$ gauge fields. 
Note that by a partial integration a CS term $(ij)$ becomes a term $(ji)$, 
so these  terms are equivalent.

Let us  also assume that this theory features a family of Dirac monopoles 
for each compact $U(1)$ gauge group. That is, the spectrum of Dirac monopoles 
consists of the magnetic charges 
$g^{(i)} = \frac{2\pi m^{(i)}}{e^{(i)}}$ with $m^{(i)} \in \Z$ 
for  $1 \leq i \leq k$. In this 2+1 dimensional Minkowsky setting
these monopoles are instantons tunneling between states with flux 
difference $\Delta \phi^{(i)}  = \Delta \int \! 
d\,^2x \,\epsilon^{kl}\partial_k A^{(i)\, l}=\frac{2\pi m^{(i)}}{e^{(i)}}$. 
It can be shown that a consistent implementation of these 
monopoles requires  
that the topological masses in~(\ref{csact}) are 
quantized as~\cite{spabcst}
\bea                                  
\mu^{(i)} \;=\; \frac{p^{(i)} e^{(i)}e^{(i)}}{\pi} 
\;\;\; \mbox{and} \;\;\; 
\mu^{(ij)} \;=\; \frac{p^{(ij)} e^{(i)}e^{(j)}}{\pi} 
\;\;\; \mbox{with $p^{(i)},
p^{(ij)} \in {\mbox{\bf Z}}$} \,.       
\label{quantmu}  
\eea
It is known that in contrast to ordinary 2+1 dimensional
QED, the presence of Dirac monopoles in these massive
gauge theories does {\em not} lead to confinement of the matter 
charges $q^{(i)}$.

The spectrum of the theory defined 
by~(\ref{alg}) with~(\ref{ymh})--(\ref{csact}) 
contains $k$ different quantized matter charges $q^{(i)}=n^{(i)}e^{(i)}$
with $n^{(i)} \in \Z$, $k$ different 
vortex species carrying quantized magnetic flux 
$\phi^{(i)} = \frac{2\pi a^{(i)}}{N^{(i)} e^{(i)}}$ with  
$a^{(i)} \in \Z$ and dyonic combinations of these charges and 
vortices. Since all gauge fields are massive, there are 
no long range Coulomb interactions between these particles. 
The remaining long range interactions are AB interactions.
As has been explained in~\cite{sam},
a counterclockwise monodromy of a vortex $\phi^{(i)}$ and a charge
$q^{(i)}$ gives rise to the AB phase $\exp(\im q^{(i)}\phi^{(i)})$ in 
the wave function. The crucial point was that the Higgs mechanism replaces 
the fluxes attached to the charges $q^{(i)}$ in the unbroken 
CS phase~\cite{schonfeld} by screening charges which screen the Coulomb 
fields around the charges but do {\em not} couple to the AB 
interactions. Hence, contrary to the unbroken CS phase there are {\em no} 
AB interactions among the charges in the CS Higgs phase. 
Instead, the CS term~$(i)$ in~(\ref{csact}) now implies the AB phase 
$\exp(\im \mu^{(i)} \phi^{(i)} \phi^{(i)'})$ for a counterclockwise 
monodromy of  two remote vortices $\phi^{(i)}$ and $\phi^{(i)'}$, 
whereas a CS term~$(ij)$ gives rise to the AB
phase $\exp(\im \mu^{(ij)} \phi^{(i)} \phi^{(j)})$ for a counterclockwise 
monodromy of two remote vortices $\phi^{(i)}$ and 
$\phi^{(j)}$ \cite{spabcst}. 
Let us label the particles in this theory as 
$\left( A,n^{(1)} \!\! \ldots  n^{(k)}\right)$ with 
$A := \left(a^{(1)},\ldots,a^{(k)} \right)$ and $a^{(i)}, n^{(i)} \in \Z$. 
Upon implementing~(\ref{quantmu}), 
the foregoing AB interactions can then be recapitulated as
\bea                             
{\cal R}^{2 \;\; A \qquad \;\;\;\;\;\;A'}_{\; \; \;n^{(1)} \ldots n^{(k)} \;\;
n^{(1)'} \ldots n^{(k)'} } &=&  
\varepsilon_A(A') \; \Gamma^{n^{(1)} \ldots n^{(k)}}(A') \; 
\varepsilon_{A'} (A) \; \Gamma^{n^{(1)'} \ldots n^{(k)'}}(A) \, .        
                    \label{brz2}
\eea 
The indices attached to the monodromy operator 
${\cal R}^2$ express the fact that it acts on the particles 
$\left( A,n^{(1)} \!\! \ldots  n^{(k)}\right)$ and 
$\left( A',n^{(1)'} \!\! \ldots  n^{(k)'}\right)$, whereas
\bea
\varepsilon_A(A') &:=& \exp \left( \sum_{1\leq i < j \leq k}
\frac{2 \pi \im p^{(i)}}{N^{(i)} N^{(i)}} 
\, a^{(i)}a^{(i)'} + 
\frac{2 \pi \im p^{(ij)}}{N^{(i)}N^{(j)}} 
\, a^{(i)}a^{(j)'}\right),  \label{epsi} 
\eea 
and $\Gamma^{n^{(1)} \! \ldots n^{(k)}} (A) 
:=   \exp \left( 
\sum_{i=1}^k \frac{2 \pi \im}{N^{(i)}} \, n^{(i)} a^{(i)} \right)$.   
It can also be shown that the particles in this theory satisfy the 
canonical spin-statistics connection:
\bea \label{spinstatis}
\exp \left(\im \Theta_{(A,n^{(1)} \! \ldots  n^{(k)})}\right) \;=\;
\exp \left(2 \pi \im s_{(A,n^{(1)} \! \ldots  n^{(k)})}\right) \;=\;
\varepsilon_A(A) \; \Gamma^{n^{(1)}\! \ldots n^{(k)}}(A) \, ,
\eea
with  $\exp (\im \Theta_{(A,n^{(1)} \! \ldots  n^{(k)})})$ 
the quantum statistics phase resulting from a counterclockwise 
braid operation ${\cal R}$ on  two identical particles
$(A,n^{(1)} \! \ldots  n^{(k)})$
and $s_{(A,n^{(1)} \! \ldots  n^{(k)})}$ the spin assigned to these 
particles.
Under the remaining (long range) 
AB interactions~(\ref{brz2}) and~(\ref{spinstatis}), 
the charge labels $n^{(i)}$ 
clearly become $\Z_{N^{(i)}}$ quantum numbers. Also, in the presence 
of the aforementioned Dirac monopoles
the fluxes $a^{(i)}$ are conserved modulo $N^{(i)}$.
Specifically, the tunneling events induced by the minimal monopoles 
read~\cite{spabcst}
\bea                             \label{instb1}
\mbox{monopole $(i)$: } \left\{   \begin{array}{lcl}
a^{(i)} & \mapsto & a^{(i)} -N^{(i)}  \\
n^{(i)} & \mapsto & n^{(i)}  +  2p^{(i)} \, , \;
n^{(j)} \;  \mapsto \;  n^{(j)}  + p^{(ij)} \, .
\end{array} \right.                                           
\eea   
Hence, the decay of an unstable flux $a^{(i)}$ through a monopole $(i)$ 
is accompanied by the creation of matter charges of species and strength
depending on the CS parameters~(\ref{quantmu}).
It is  easily verified that these {\em local} tunneling events are invisible 
to the long range AB interactions~(\ref{brz2}) and that the particles 
connected by the monopoles have the same spin factor~(\ref{spinstatis}).
So the spectrum of this theory compactifies to
$ 
\left( A,n^{(1)} \!\! \ldots  n^{(k)}\right)$  
with $A=\left(a^{(1)}, \ldots, a^{(k)} \right)$ and  $a^{(i)}, n^{(i)}
\in 0,1, \ldots, N^{(i)}-1$,           
where the modulo calculus for the flux quantum numbers $a^{(i)}$ 
involves the charge jumps in~(\ref{instb1}).
Moreover, it can be shown that the CS parameters 
$p^{(i)}$ and  $p^{(ij)}$
become periodic with period $N^{(i)}$ and 
greatest common divisor ${\gcd}(N^{(i)},N^{(j)})$ 
of $N^{(i)}$ and $N^{(j)}$ respectively~\cite{spabcst}. 
That is, up to a relabeling of the dyons, the broken CS theory defined by 
$p^{(i)}$ and  $p^{(ij)}$
describes the same spectrum and AB interactions as that defined by 
$p^{(i)}+N^{(i)}$ and 
$p^{(ij)}+{\gcd}(N^{(i)},N^{(j)})$. 
Finally, note that the additional AB phases~(\ref{epsi}) among the  
vortices and the twisted tunneling properties of the 
monopoles~(\ref{instb1}) form the only distinction with the abelian discrete 
gauge theory describing the long distance physics in the absence 
of the CS action~(\ref{csact}). As will be explained 
in the following sections, this distinction is completely encoded in a 
3-cocycle for the residual gauge group 
$\Z_{N^{(1)}} \times \cdots \times \Z_{N^{(k)}} $.

\section{Group cohomology and symmetry breaking}
\label{bgt}

\noindent
A deep result due to Dijkgraaf and Witten~\cite{diwi}
states that the  CS actions $S_{\mbox{\scriptsize CS}}$ for 
a compact gauge group $G$ are in one--to--one correspondence
with the elements of the cohomology 
group $H^4 (B{ G}, \Z)$ of the classifying space
$B{ G}$ with integer coefficients $\Z$.
In particular, this classification includes the case of finite groups $H$. 
The isomorphism $H^4 ({BH}, \Z)  \simeq  H^3 ({ H}, U(1))$
which is only valid for finite ${ H}$ then implies  
that the different CS theories 
for a finite gauge group $H$ correspond to the different 
elements $\omega \in H^3({ H}, U(1))$, i.e.\ algebraic 
3-cocycles $\omega$ taking values in $U(1)$.

One of the new observations of~\cite{spabcst}
was that the long distance physics of the spontaneously 
broken model~(\ref{alg}) is described by a CS theory with finite gauge 
group $H$ and 3-cocycle $\omega \in H^3({ H}, U(1))$ determined 
by the original CS action $S_{\mbox{\scriptsize CS}} \in H^4 (B{ G}, \Z)$ 
for the broken gauge group $G$ by the natural homomorphism 
\bea                          
H^4 (B{ G}, \Z)  &\longrightarrow&   H^3 ({ H}, U(1)) \, , 
\label{homo}   
\eea 
induced by the inclusion $H \subset G$.
The physical picture behind this homomorphism, also known as
the restriction, is that the CS term 
$S_{\mbox{\scriptsize CS}}$ gives rise to additional AB interactions
among the magnetic vortices which are completely encoded
in the 3-cocycle $\omega$ for the finite residual gauge group $H$
being the image of  $S_{\mbox{\scriptsize CS}}$ under the homomorphism 
(\ref{homo}).

Let me illustrate these general remarks with the abelian 
example of the previous section where $G\simeq U(1)^k$ and 
$H \simeq \Z_{N^{(1)}} \times \cdots \times
\Z_{N^{(k)}} $. A simple calculation~\cite{spabcst} shows 
$
H^4(B(U(1)^k), \Z) \simeq \Z^{ k + \frac{1}{2}k(k-1)}. 
$
Note that this classification of the CS actions for the compact 
gauge group $G\simeq U(1)^k$ is indeed in agreement 
with~(\ref{csact})--(\ref{quantmu}), i.e.\ the integral 
CS parameters $p^{(i)}$ and $p^{(ij)}$ provide a complete labeling
of  the elements of $H^4(B(U(1)^k), \Z)$. 
To proceed, it can be shown~\cite{spabcst} that for 
$H \simeq \Z_{N^{(1)}} \times \cdots \times\Z_{N^{(k)}} $
\bea  \label{conj3e}
H^3(H,U(1)) 
&\simeq& \! \bigoplus_{1 \leq i < j < l \leq k} \! \Z_{N^{(i)}}
         \oplus \Z_{{\gcd} (N^{(i)},N^{(j)})}
         \oplus \Z_{{\gcd} (N^{(i)},N^{(j)},N^{(l)})} \, . 
\eea
Let $A,B$ and $C$ denote elements of $\Z_{N^{(1)}} \times \cdots \times
\Z_{N^{(k)}}$, so
$
A := \left( a^{(1)} , a^{(2)}, 
\ldots, a^{(k)} \right)$ with 
$a^{(i)} \in \Z_{N^{(i)}}$   
for $i=1,\ldots, k$
and similar decompositions for  $B$ and $C$. I adopt the additive 
presentation, i.e.\ 
the elements $a^{(i)}$ of $\Z_{N^{(i)}}$ take values in the range 
$ 0,\ldots,N^{(i)}-1$ and group multiplication is defined as:
$
A \cdot B  =  [A+B]  :=  \left([a^{(1)}+b^{(1)}],  \ldots ,
[a^{(k)}+b^{(k)}]\right)$.
The rectangular brackets denote modulo $N^{(i)}$ 
calculus such that the sum always lies in the range $0, \ldots, N^{(i)}-1$.
The most general 3-cocycle
for $\Z_{N^{(1)}} \times \cdots \times \Z_{N^{(k)}}$ can then be presented 
as some product of
\bea 
\omega^{(i)}(A,B,C) &=& 
\exp \left( \frac{2 \pi \im p^{(i)}}{N^{(i)\;2}} \;
a^{(i)}\left(b^{(i)} +c^{(i)} -[b^{(i)}+c^{(i)}]\right) \right) 
\label{type1}    \\
\omega^{(ij)}(A,B,C) &=& 
\exp \left( 
\frac{2 \pi \im p^{(ij)}}{N^{(i)}N^{(j)}}  \;
a^{(i)}\left(b^{(j)} +c^{(j)} - [b^{(j)}+c^{(j)}]\right) \right)
   \label{type2}  \\
\omega^{(ijl)}(A,B,C)  &=& 
\exp \left( \frac{2 \pi \im
p^{(ijl)}}{{\gcd}(N^{(i)}, N^{(j)},N^{(l)})} \;
a^{(i)}b^{(j)}c^{(l)} \right) , \label{type3}
\eea
with $1 \leq i < j < l \leq k$. The integral parameters 
$p^{(i)}$, $p^{(ij)}$ and 
$p^{(ijl)}$  label the different 
elements of~(\ref{conj3e}). It can be verified that in agreement 
with~(\ref{conj3e}) the functions~(\ref{type1}), (\ref{type2}) 
and (\ref{type3}) are periodic in these parameters with period $N^{(i)}$, 
${\gcd}(N^{(i)},N^{(j)})$ and ${\gcd}(N^{(i)},N^{(j)},N^{(l)})$ respectively. 
It is also readily checked that these three functions 
and their products indeed satisfy the 3-cocycle relation
\bea
\label{pentagon}
\delta\omega(A,B,C,D) \; = \;
\frac{\omega(A,B,C)\;\omega(A,B \cdot C,D)\;\omega(B,C,D)}{ 
\omega(A \cdot B,C,D)\;\omega(A,B,C \cdot D)} \; = \; 1 \, ,  
\eea
where $\delta$ denotes the coboundary operator.

We are now ready to make the 
homomorphism~(\ref{homo}) accompanying the spontaneous breakdown 
of the gauge group $U(1)^k$ to  
$\Z_{N^{(1)}} \times \cdots \times \Z_{N^{(k)}}$ explicit.
In terms of the integral CS parameters~(\ref{quantmu}), 
it takes the form~\cite{spabcst}
\bea
H^4(B(U(1)^k), \Z) 
&\longrightarrow & 
H^3(\Z_{N^{(1)}} \times \cdots \times \Z_{N^{(k)}}, U(1)) 
\label{homo1en2} \\
p^{(i)} & 
\longmapsto & p^{(i)} \qquad \; \bmod  N^{(i)} 
\label{homoI} \\
p^{(ij)} & \longmapsto & 
p^{(ij)} \qquad \bmod  
\gcd(N^{(i)}, N^{(j)}) \, .     \label{homoII}
\eea
The periodic parameters being the images of this mapping label the 
3-cocycles~(\ref{type1}) and~(\ref{type2}). So 
the long distance physics of the spontaneously broken 
$U(1)^k$ CS theory~(\ref{alg})--(\ref{quantmu}) is described by 
a $\Z_{N^{(1)}} \times \cdots \times \Z_{N^{(k)}}$ CS theory with 
3-cocycle being the product 
$\omega=\prod_{1\leq i < j \leq k} \omega^{(i)}
\omega^{(ij)}$. That this 3-cocycle indeed 
leads to the additional AB phases~(\ref{epsi}) and 
the twisted tunneling properties~(\ref{instb1}) will become clear 
in the next section. Finally, 
note that the image of~(\ref{homo1en2}) does not 
contain the 3-cocycles~(\ref{type3}). In other words,
abelian discrete CS theories defined by 
these 3-cocycles can not be obtained from 
the  spontaneous  breakdown of a  $U(1)^k$ CS theory.

\section{The quasi-quantum double}

\noindent
The quasi-quantum double $\DW$ related to a CS theory 
with finite abelian gauge group $H \simeq 
\Z_{N^{(1)}} \times \cdots \times \Z_{N^{(k)}}$ and some 3-cocycle
$\omega$ is spanned by the elements $\{ {\mbox{P}}_A \, B \}_{A,B\in { H}} $
representing a global symmetry transformation $B \in H$ followed
by the operator ${\mbox{P}}_A$ projecting 
out the magnetic flux $A\in H$. In this basis, multiplication, 
and comultiplication are defined as~\cite{dpr,spm1,spabcst}
\bea
{\mbox{P}}_A \, B \cdot {\mbox{P}}_D \, C &=& 
\delta_{A,D} \;\; {\mbox{P}}_A  \, B \cdot C 
\;\; c_A(B,C) \label{algebra}       \\
 \Delta(\,{\mbox{P}}_A \, B \,) &=& 
\sum_{C\cdot D=A} \; {\mbox{P}}_C \, B \ot {\mbox{P}}_D \, B       \;\;
c_B(C,D) \, , \label{coalgebra} 
\eea
with $c_A(B,C) := \frac{\omega (A,B,C) \omega(B,C,A)}{\omega(B,A,C)}$ and
$\delta_{A,B}$ the Kronecker delta.
From~(\ref{pentagon}) it  follows   
that $c_A$ satisfies the 2-cocycle relation
$\delta c_A(B,C,D)=  \frac{c_A(B,C)  c_A(B \cdot C, D)}{c_A(B,C \cdot D) 
 c_A(C,D)}=1$, which implies that the multiplication~(\ref{algebra})
is associative and that the comultiplication~(\ref{coalgebra}) 
is quasi-coassociative:
$
({\mbox{id}} \ot \Delta) \, \Delta( \, {\mbox{P}}_A \, B \, ) = 
\varphi\cdot (\Delta \ot {\mbox{id}}) \, \Delta( \, {\mbox{P}}_A \, B \, ) 
\cdot\varphi^{-1} 
$ with $\varphi := \sum_{A,B,C}\,\omega^{-1}(A,B,C) \;
{\mbox{P}}_A \otimes {\mbox{P}}_B \otimes {\mbox{P}}_{C}$.

The different particles in the associated CS theory are  
labeled by their magnetic flux $A \in H$ paired with a projective  
UIR $\alpha$ of $H$ defined as ${\alpha}(B) \cdot  {\alpha}(C) = c_A(B,C)
 \;  {\alpha}(B \cdot C)$.
Each particle $( \, A, {\alpha} \,)$ is equipped with an internal Hilbert 
space $V_{\alpha}^A$ (spanned by the states
$
 \{|\, A,\,^{\alpha} v_j\rangle\}_{j=1,\ldots,d_\alpha}  
$
with $^{\alpha}\!v_j$ a basis vector and $d_\alpha$ the dimension of 
$\alpha$) carrying an irreducible 
representation $\Pi^A_{\alpha}$ 
of $D^{\omega}({ H})$ given by~\cite{dpr} 
\bea \label{13}                                              
\Pi^A_{\alpha}(\, {\mbox{P}}_B \, C \,) \; |\,A ,\,^{\alpha} v_j \rangle &=&
\delta_{A,B}\;\, |\,A,\,\alpha(C)_{ij}\,^{\alpha} v_i \rangle \, .
\eea

In the process of rotating a particle $(A,\alpha)$
over an angle of $2 \pi$ its charge $\alpha$ 
is transported around the flux $A$ and as a result picks 
up a global transformation $\alpha(A)$.
With~(\ref{13}) it is easily checked that this AB effect is implemented
by the central element $\sum_B \; {\mbox{P}}_B \, B$. 
Schur's lemma then implies:   
$
\alpha(A) = e^{2 \pi \im s_{(A,\alpha)}} \; {\mbox{\bf 1}}_\alpha $
with $s_{(A,\alpha)}$ the spin assigned to the particle 
$( \, A, {\alpha} \,)$ and ${\mbox{\bf 1}}_\alpha$
the unit matrix.

The action~(\ref{13}) of $\DW$ is extended to two-particle 
states by means of the comultiplication~(\ref{coalgebra}).
Specifically, the  
representation $(\Pi^A_{\alpha} \ot \Pi^B_{\beta}, V_{\alpha}^A \ot 
V_{\beta}^B)$ of $\DW$ related to a system of two particles 
$(\,A, \alpha \,)$ and $(\,B, \beta \,)$ is defined by the action
$\Pi^A_{\alpha} \ot \Pi^B_{\beta}( \Delta (\, {\mbox{P}}_A \, B \, ))$. 
The tensor product representation of  $\DW$ related to
a system of three particles $(\,A, \alpha \,)$, $(\,B, \beta \,)$ and 
$(\,C, \gamma \,)$ may now be constructed
through $(\Delta \ot {\mbox{id}} ) \, \Delta$
or through $({\mbox{id}} \ot \Delta) \, \Delta$.
The aforementioned quasi-coassociativity relation implies that 
these two constructions are equivalent.

The braid operator ${\cal R}$ establishing  
a counterclockwise interchange of two particles $(\, A, \alpha \,)$
and $(\, B, \beta \,)$ is defined as 
\bea                 \label{braidaction}
{\cal R} \;| \, A,\, ^{\alpha} v_j\rangle
|\, B,\,^{\beta} v_l\rangle &=& 
|\,B,\,{\beta}(A)_{ml} \, ^{\beta} v_m\rangle 
|\,A,\, ^{\alpha} v_j\rangle \, .
\eea

The tensor product representation 
$(\Pi^A_{\alpha} \ot \Pi^B_{\beta},V_\alpha^A \ot V_\beta^B)$ 
in general decomposes into a direct sum of irreducible 
representations $(\Pi^C_{\gamma},V_\gamma^C)$
\bea               \label{piet}
\Pi^A_{\al}\otimes\Pi^B_{\beta}& = & \bigoplus_{C , \gamma}
N^{AB\gamma}_{\alpha\beta C} \; \Pi^C_{\gamma} \, .
\eea
This so-called fusion rule determines which particles 
$(\,C,\gamma \, )$ can be formed in the composition 
of two particles $(\, A,\alpha \,)$ and  $(\, B,\beta\,)$.
The modular matrices $S$ and $T$ associated to 
the fusion algebra~(\ref{piet}) are determined by the braid 
operator~(\ref{braidaction}) and the spin factors 
$e^{2\pi \im s_{(A,\alpha)}}:=\alpha(A)/d_\alpha$
\bea                                
S^{AB}_{\alpha\beta} \; := \; \frac{1}{|{ H}|} \, \mbox{tr} \; {\cal 
R}^{-2 \; AB}_{\; \; \; \; \; \alpha\beta} \qquad \mbox{and} \qquad
T^{AB}_{\alpha\beta} \; := \; 
\delta_{\alpha,\beta} \, \delta^{A,B} \; 
\exp\left(2\pi \im s_{(A,\alpha)}\right) \, .\label{modular}
\eea
$|H|$ denotes the order of $H$ and tr abbreviates trace.
As usual, the multiplicities in~(\ref{piet}) can be expressed in 
terms of the matrix $S$ by means of Verlinde's formula:
\bea      \label{verlindez}
N^{AB\gamma}_{\alpha\beta C}&=&\sum_{D,\delta}\frac{
S^{AD}_{\alpha\delta}S^{BD}_{\beta\delta}
(S^{*})^{CD}_{\gamma\delta}}{S^{eD}_{0\delta}} \, .
\eea

\begin{figure}[htb]    \epsfxsize=11cm
\centerline{\epsffile{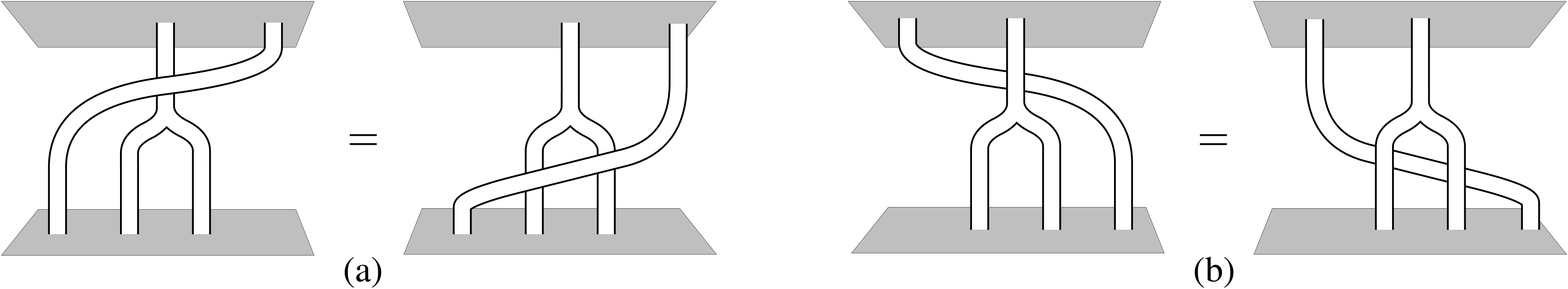}}
\fcaption{The diagrams in (a) and (b) (with the ribbons representing 
particle--trajectories) are homotopic. So the 
result of braiding a particle with 
two particles separately or with the composite that arises after fusion should 
be the same.}
\label{qu1zon}
\end{figure}
\begin{figure}[htb]    \epsfxsize=10cm
\centerline{\epsffile{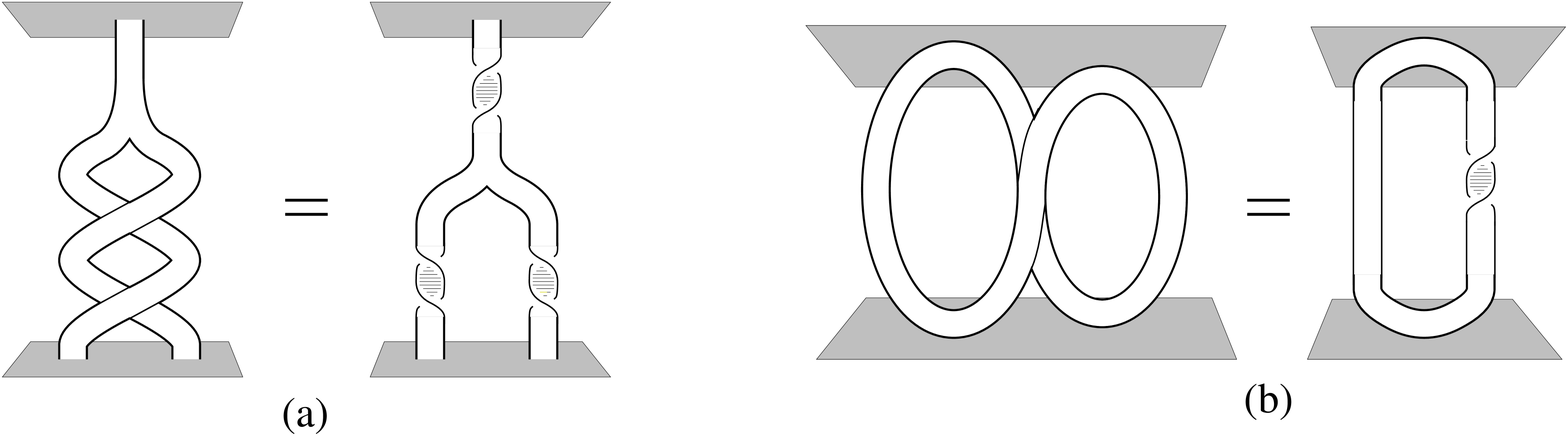}}
\fcaption{The fact that the ribbon diagrams in (a) are homotopic
indicates that the result of 
a counteclockwise monodromy of two particles in a given fusion channel 
followed  by fusion of the pair should be the same as a clockwise
rotation of the two particles seperately followed 
by fusion of the pair and a counterclockwise rotation 
of the composite. The fact that the diagrams in (b) are homotopic implies 
that the effect of a counterclockwise interchange of two particles in two
identical but separate particle/anti-particle pairs should be 
the same as a counterclockwise rotation of a (anti-)particle in 
one of these pairs.}
\label{kanaal}
\end{figure}

It is impossible to do justice to the complete 
structure of $\DW$ in this limited number of pages. 
A detailed treatment can be found in~\cite{spabcst}.
Let me just flash some pictures giving an impression of some 
relations unmentioned so far. First of all, the 
comultiplication~(\ref{coalgebra}), braid operator~(\ref{braidaction}) and 
associator $\varphi$ obey
the so-called quasitriangularity equations expressing  
the compatibility of fusion and braiding depicted in Fig.~\ref{qu1zon}.
As an immediate consequence the braid operators 
satisfy the quasi--Yang--Baxter equation implying that the multi-particle
internal Hilbert spaces carry a (possibly reducible)
representation of the braid group.
Since the braid operators~(\ref{braidaction}) are of finite order,
the more accurate statement is that we are dealing with representations
of truncated braid groups being factor groups of the braid 
group~\cite{banff,spabcst}.  The quasi-triangularity equations also state that
the action of the truncated braid group commutes with the action of $\DW$.
Thus the multi-particle internal Hilbert spaces, in fact, carry
a (possibly reducible) representation of the direct product 
of $\DW$ and some truncated braid group. Further, to keep track of 
the writhing of the particle--trajectories and the resulting  
spin factors these are represented by 
ribbons. Passing from worldlines to `worldribbons' can only be consistent
if the demands in Fig.~\ref{kanaal} are met. The consistency demand in
Fig.~\ref{kanaal}(a) is met by the generalized spin-statistics 
connection $K^{ABC}_{\alpha\beta\gamma} 
{\cal R}^2 = 
e^{2\pi \im(s_{(C,\gamma)}-s_{(A,\alpha)}-s_{(B,\beta)})} 
K^{ABC}_{\alpha\beta\gamma}$ with $K^{ABC}_{\alpha\beta\gamma}$ 
the projection on the channel $(C,\gamma)$  in~(\ref{piet})
and the demand in  Fig.~\ref{kanaal}(b) by the canonical spin-statistics 
connection $K^{AAC}_{\alpha\alpha\gamma} 
{\cal R} = 
e^{2\pi \im s_{(A,\alpha)}} 
K^{AAC}_{\alpha\alpha\gamma}$ which only holds for the fusion channels
$(C,\gamma)$ in which both particles $(A,\alpha)$ are in identical
internal quantum states.

Let me finally  establish that the CS term~(\ref{csact})
in the broken model of section~2 indeed boils down to the 
3-cocycle $\omega=\prod_{1\leq i < j \leq k} \omega^{(i)}
\omega^{(ij)}$ as indicated by~(\ref{homo1en2}). To start with, it is readily 
checked that the 2-cocycle $c_A$ entering~(\ref{algebra}) 
and~(\ref{coalgebra}) for this $\omega$  is trivial, i.e.\ 
$
c_A(B,C)  =  \delta \varepsilon_A (B,C)  = 
\frac{\varepsilon_A(B)  \varepsilon_A(C)}{\varepsilon_A(B \cdot C)} 
$
with $\varepsilon_A$  given by~(\ref{epsi}).
Thus the dyon charges $\alpha$ in~(\ref{piet}) for this 
$\omega$ are trivial projective representations 
of $H$ of the form $\alpha(C) = \varepsilon_A(C) 
\Gamma^{n^{(1)} \! \ldots n^{(k)}}(C)$ 
with $\Gamma^{n^{(1)} \! \ldots n^{(k)}}$ the ordinary UIR of 
$H$ appearing in~(\ref{brz2}).
It is now easily verified that the monodromy operator following 
from~(\ref{braidaction}) for this case is the same as~(\ref{brz2}),
that the spinfactors $e^{2\pi \im s_{(A,\alpha)}}=\alpha(A)$ 
coincide with~(\ref{spinstatis}) and 
that the fusion rules~(\ref{piet}) following from~(\ref{modular}) 
and~(\ref{verlindez}) reproduce the tunneling 
properties~(\ref{instb1}) of the Dirac monopoles. \qed

\section{Nonabelian electric/magnetic dualities}

\noindent 
The 3-cocycles~(\ref{type3}) that can not be reached from a 
spontaneously broken $U(1)^k$ CS theory are actually the most interesting.
They  render an {\em abelian} discrete $H$ gauge theory 
{\em nonabelian} and generally lead to dualities with 2+1 dimensional
theories with a nonabelian finite gauge group of the same order as $H$. 
The point is that the 2-cocycles $c_A$ appearing 
in~(\ref{algebra}) and~(\ref{coalgebra}) 
for such a 3-cocycle $\omega^{(ijl)}$ are {\em nontrivial}, so the dyon 
charges $\alpha$ in~(\ref{piet}) become nontrivial (i.e.\ {\em higher} 
dimensional) projective UIR's of $H$.
Consequently, the braid operator~(\ref{braidaction}) now generally 
acts as a matrix leading to the usual host of nonabelian 
phenomena~\cite{banff,spabcst}
such as nonabelian braid statistics, nonabelian AB scattering, exchange
of nontrivial Cheshire charges and quantum statistics 
between particle pairs through monodromy processes, etc...

Let me briefly illustrate these general remarks with the simplest 
example~\cite{spabcst},  namely a CS theory with gauge group 
${ H} \simeq \Z_2^3 := 
\Z_2 \times \Z_2  \times \Z_2 $ defined by the corresponding 
nontrivial 3-cocycle~(\ref{type3}). 
Whereas the ordinary $\Z_2^3$ theory
features 64 different singlet particles, it turns out that the spectrum 
just consists of 22 different particles in the presence of this 
3-cocycle $\omega^{(123)}$. Specifically, the dyon charges 
which formed 1-dimensional UIR's of $\Z_2^3$
are  reorganized into 2-dimensional or doublet projective 
UIR's  of $\Z_2^3$, i.e.\ besides the 
8 singlets (the vacuum and 7 ordinary nontrivial 
$\Z_2^3$ charges) the spectrum now contains 
14 doublet dyons carrying a nontrivial
(singlet) flux and a nontrivial projective $\Z_2^3$ doublet charge.
Further, there are only two nonabelian finite groups of order 
$|\Z_2^3|=8$: 
the dihedral group $D_4$ and the double dihedral  
group $\bar{D}_2$. Like the $\{\Z_2^3 , \omega^{(123)}\}$ CS theory, the 
spectrum of the theories with gauge group $D_4$
and $\bar{D}_2$ both contain 8 singlet particles and 14 doublet particles  
albeit of different nature.  It can be checked that the 
$\{\Z_2^3 , \omega^{(123)}\}$ CS theory 
is dual to the $D_4$ theory, i.e.\ the exchange 
$\{\Z_2^3 , \omega^{(123)}\}\leftrightarrow D_4$ corresponds to an invariance 
of the modular matrices~(\ref{modular}) indicating that
these two theories indeed describe the same spectrum {\em and} the same
topological interactions. The duality transformation 
exchanges the projective dyon charges in the 
$\{\Z_2^3 , \omega^{(123)}\}$ CS theory with the magnetic doublet fluxes 
in the $D_4$ theory. So we are actually dealing with
some kind of nonabelian electric/magnetic duality.   
Let me also note that  adding a 3-cocycle~(\ref{type2}) does not 
spoil this duality, i.e.\ we also have the dualities  
$\{\Z_2^3 , \omega^{(123)} \omega^{(ij)} \}
\leftrightarrow D_4$ with $1\leq i <j \leq k$.
Finally,  the 
$\{\Z_2^3 , \omega^{(123)} \omega^{(i)}\}$ 
CS theories (with $\omega^{(i)}$ given in~(\ref{type1}) and $i=1,2$ or $3$)
turn out to be  dual to the $\bar{D}_2$ theory.

\section{Concluding remarks}
\noindent
Whether the interesting 
3-cocycles~(\ref{type3}) can be reached from the spontaneous breakdown
of a nonabelian CS theory is under investigation. I am 
currently also working on the generalization of the dualities described
in the previous section to abelian finite gauge groups of order higher
then $|\Z_2^3|=8$. 
Finally, for a (concise)
discussion of CS theories in which a nonabelian compact gauge group is
spontaneously broken to a {\em nonabelian} finite subgroup, the reader 
is referred to Ref.~\cite{thesis}.
         
\nonumsection{Acknowledgements}
\noindent
I would like to thank the organizers for an inspiring workshop. This
work was partly supported by an EC grant (contract no. ERBCHBGCT940752).

\nonumsection{References}

\end{document}

\end{thebibliography}
\end{document}